\documentclass[dvips]{article}
\usepackage{icrctc07}

\title{Propagation of Ultra-high-energy Protons in Cosmic Magnetic Fields}
\shorttitle{Propagation of Ultra-high-energy Protons}
\authors{Hajime Takami$^{1}$, and Katsuhiko Sato$^{2}$}
\shortauthors{Takami \& Sato}
\afiliations{$^1$Department of Physics, School of Science, the University of Tokyo, 7-3-1, Hongo, Bunkyo-ku, Tokyo, 113-0033, Japan\\ $^2$Research Center for the Early Universe, School of Science, the University of Tokyo, 7-3-1, Hongo, Bunkyo-ku, Tokyo, 113-0033, Japan}
\email{takami@utap.phys.s.u-tokyo.ac.jp}

\abstract{We simulate the arrival distribution of 
ultra-high-energy (UHE) protons 
by following their propagation processes 
in several strengths of a structured extragalactic magnetic field (EGMF). 
Comparing our result to observational one by Akeno Giant Air Shower Array, 
we constrain the number density of UHE cosmic ray 
sources with the small-scale anisotropy.  
As a result, 
the source number density is $\sim 10^{-5}~{\rm Mpc}^{-3}$ 
with uncertainty of about an order of magnitude 
due to the small number of observed events. 
This hardly depends on our structured EGMF strength. 
We also investigate future prospects for this approarch. 
The near future observations, such as Pierre Auger Observatory, 
can distinguish $10^{-6}~{\rm Mpc}^{-3}$ accurately 
from the more source densities. 
Number of events to discriminate between $10^{-4}$ 
and $10^{-5}~{\rm Mpc}^{-3}$ is dependent on the EGMF strength.}

\begin{document}
\maketitle

\section{Introduction}

The origin of ultra-high-energy cosmic rays (UHECRs) is 
one of challenging problems in astroparticle physics. 
One of significant information on UHECR sources is 
their arrival distribution. 
Akeno Giant Air Shower Array (AGASA) reports 
small-scale anisotropy (SSA) within a few degree scale 
while large-scale isotropy (LSI) based on a harmonic analysis 
\footnote{Such AGASA results conflict with 
High Resolution Fly's eye(HiRes) reports, 
which finds no significant SSA \cite{abbasi04}. 
However, this discrepancy is not statistically significant 
due to the small number of observed event \cite{yoshiguchi04}.}
\cite{takeda99}. 

The SSA is predicted if UHECRs are of astrophysical origin 
with very small number. 
Therefore, the SSA has constrained their source number density 
to $10^{-5}~{\rm Mpc}$ for no extragalactic magnetic field (EGMF) 
\cite{blasi04,kachelriess05} and $10^{-6}~{\rm Mpc}$ 
for a simply uniform turbulent EGMF\cite{yoshiguchi03}. 

Recent simulations of cosmological structure formation 
predict structured magnetic fields \cite{sigl04,dolag05} 
which roughly trace the baryon density distribution. 
Tuhs, the EGMF is also structured. 
In last year, we discussed the propagation of UHE protons in a structured EGMF 
that reproduces the observed local structure \cite{takami06}, 
and the arrival distribution. 
However, we discussed only one EGMF strength, 
which is normalized to $0.4\mu{\rm G}$ at the center of the Virgo cluster. 
Observations of magnetic fields in a cluster have large uncertainty 
in the range of $0.1$- a few $\mu$G \cite{vallee04}. 
Thus, it is very important to investigate the propagation 
and constraints on UHECR sources in several strengths of the EGMF. 

In this study, 
we discuss the propagation of UHE protons in several strengths 
of the EGMF and a Galactic magnetic field (GMF), 
and compare the resulting arrival distributions 
with observational results by AGASA. 
From the comparison, 
the number density of UHECR sources is constrained. 
Such constraint has large uncertainty 
due to the the small number of observed events at present. 
So, we also discuss the possibility of a decrease in the uncertainty 
with future observations. 

\section{Numerical Methods \& Model}

The propagation of UHE protons is calculated by an application 
of the backtracking method, which is a method 
developed in our paper \cite{takami06}. 
It is very insufficient to calculate their propagation forward 
since cosmic rays do not always reach the Earth under finite EGMF 
even if they are injected toward the Earth from their sources. 

Our models of the source distribution and a structured EGMF 
are constructed out of the {\it Infrared Astronomical Satellite} 
Point Source Catalogue Redshift Survey({\it IRAS} PSCz) 
catalog of galaxies\cite{saunders00}. 
This catalog has very large sky coverage (about 84\% of all the sky). 
Thus, our source distribution and EGMF structure reflect 
large scale structures actually observed within 100 Mpc. 
The EGMF strength is normalized at the center of the Virgo cluster 
to $0.0, 0.1, 0.4$ and $1.0~{\mu}$G. 
Outside 100 Mpc, 
EGMF is assumed to be an uniform turbulence with 1nG and 
the source distribution is isotropic.
More details are written in reference \cite{takami06}. 

In this study, we adopt only a source model 
that all sources have the same power. 
In conclusion, 
we discuss results from another simple source model, 
which power of each source is proportional to its luminosity.

\section{Results}

As written above, the arrival distribution has important information 
on UHECR sources. 
We investigate the number density of UHECR sources 
which can best reproduce the AGASA results. 
Our calculated arrival distributions are compared 
to the observed arrival distribution with the two-point correlation function 
\begin{equation}
N(\theta) = \frac{1}{2\pi |\cos \theta - \cos (\theta + \Delta \theta)|} 
\sum_{\theta \leq \phi \leq \theta+\Delta \theta} 1 {\rm [sr^{-1}]}, 
\end{equation}
which is an indicator of SSA. 
Number of cosmic ray events is set to 49 events 
in the energy range of $4 \times 10^{19} < E < 10^{20}~{\rm eV}$. 
For the comparison, we define $\chi_{\theta_{\rm max}}$ as 
\begin{equation}
\chi_{\theta_{\rm max}} = \frac{1}{\theta_{\rm max}} 
\sqrt{ \sum_{\theta=0}^{\theta_{\rm max}} 
\frac{\left[ N(\theta) - N_{\rm obs}(\theta) \right]^2}
{{\sigma(\theta)}^2}}, 
\end{equation}
where $N(\theta)$ is the two-point correlation function and 
$\sigma(\theta)$ is 1$\sigma$ error due to the finite number of events. 
Small $\chi_{10}$ provides good agreement with the observational result. 

\begin{figure}
\begin{center}
\includegraphics [width=0.48\textwidth]{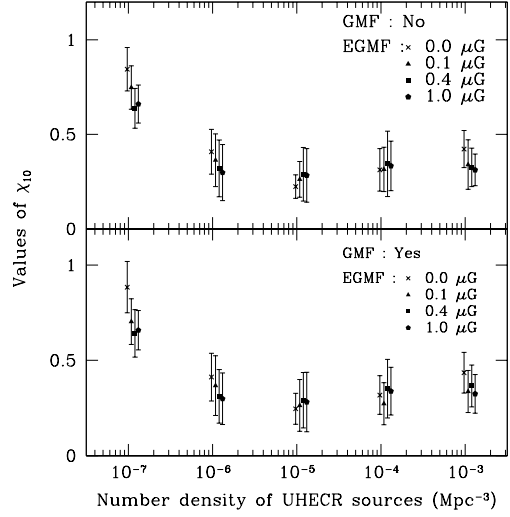}
\end{center}
\caption{$\chi_{10}$s as a function of the source number density. 
The error bars originate from 100 times source selection. 
The GMF is considered in the lower panel while not in the upper panel.}
\label{fig1}
\end{figure}

$\chi_{10}$s are shown in figure \ref{fig1}. 
While the source number density with $10^{-7}~{\rm Mpc}^{-3}$ results in 
larger value, 
the others are consistent with each other within 1$\sigma$ statistical error. 
Hence, only the SSA cannot constrain the source number density sufficiently. 

\begin{figure*}[th]
\begin{center}
\includegraphics [width=0.70\textwidth]{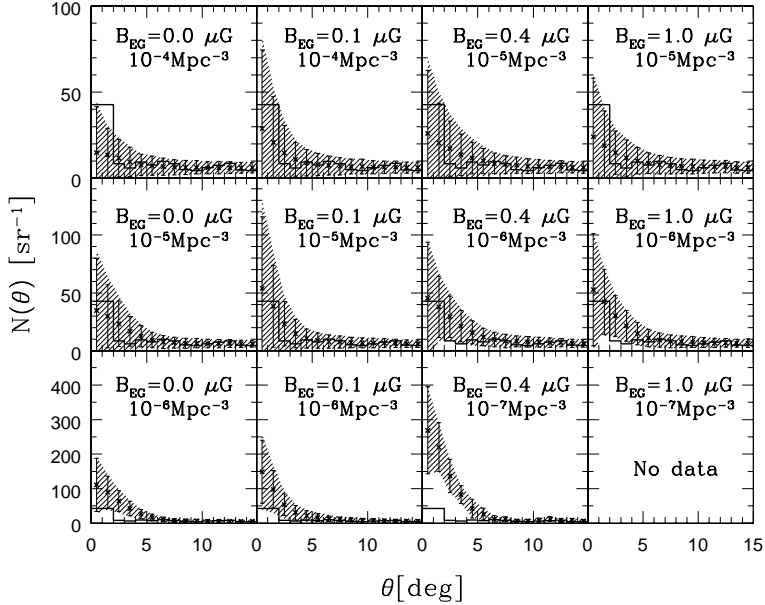}
\end{center}
\caption{The two-point correlation functions calculated from only 
source distributions that can predict the LSI. 
The histograms are the AGASA result within 
$4 \times 10^{19} < E < 10^{20}~{\rm eV}$ (49 events). 
The error bars are from the event selection for the finite events 
and the shaded regions show total 1$\sigma$ statistical errors. 
The GMF is included.}
\label{fig2}
\end{figure*}

The arrival distribution must also satisfy the LSI. 
We calculate the two-point correlation function again, 
but from merely source distribution to be able to predict LSI 
observed by AGASA. 
This is figure \ref{fig2}. 

In figure \ref{fig2}, 
the middle panels are the number densities that best reproduce AGASA results. 
However, source number densities an order of magnitude more than 
those of the best fit are consistent with the observation 
within 1$\sigma$ error except $B=0.0\mu{\rm G}$. 
On the other hand, almost all of source distributions 
with $10^{-7}~{\rm Mpc}^{-3}$ cannot satisfy the LSI. 
For $B=1.0\mu{\rm G}$, there is no source distribution 
in 100 source distributions. 
This fact can be understood in figure \ref{fig1}. 
Thus, the source number density that can best reproduce the AGASA result 
is $10^{-4} \sim 10^{-5}~{\rm Mpc}^{-3}$ for $B=0.0, 0.1\mu{\rm G}$, and 
$10^{-5} \sim 10^{-6}~{\rm Mpc}^{-3}$ for $B=0.4, 1.0\mu{\rm G}$ 
with uncertainty of about one order of magnitude. 
The source density hardly depends on EGMF strength 
since 95\% of space within 100 Mpc has not magnetic field.

The SSA and the LSI 
enable us to constrain the source number density. 
However, it has large uncertainty 
which originates probably from the small number of observed events. 
Therefore, one of our next interests is how small the uncertainty becomes 
at the Auger era. 

In order to investigate this, 
it is necessary to compare our arrival distributions with 
future observational results, which, of course, cannot be known. 
In this study, an isotropic arrival distribution is adopted 
as a template for the future results. 
If UHECR sources are of astrophysical origin and 
have a small number density, the SSA becomes stronger. 
In this viewpoint, we compare our results of the simulation 
to an isotropic distribution, using the two-point correlation function.

\begin{figure*}[th]
\begin{center}
\includegraphics[width=0.30\linewidth]{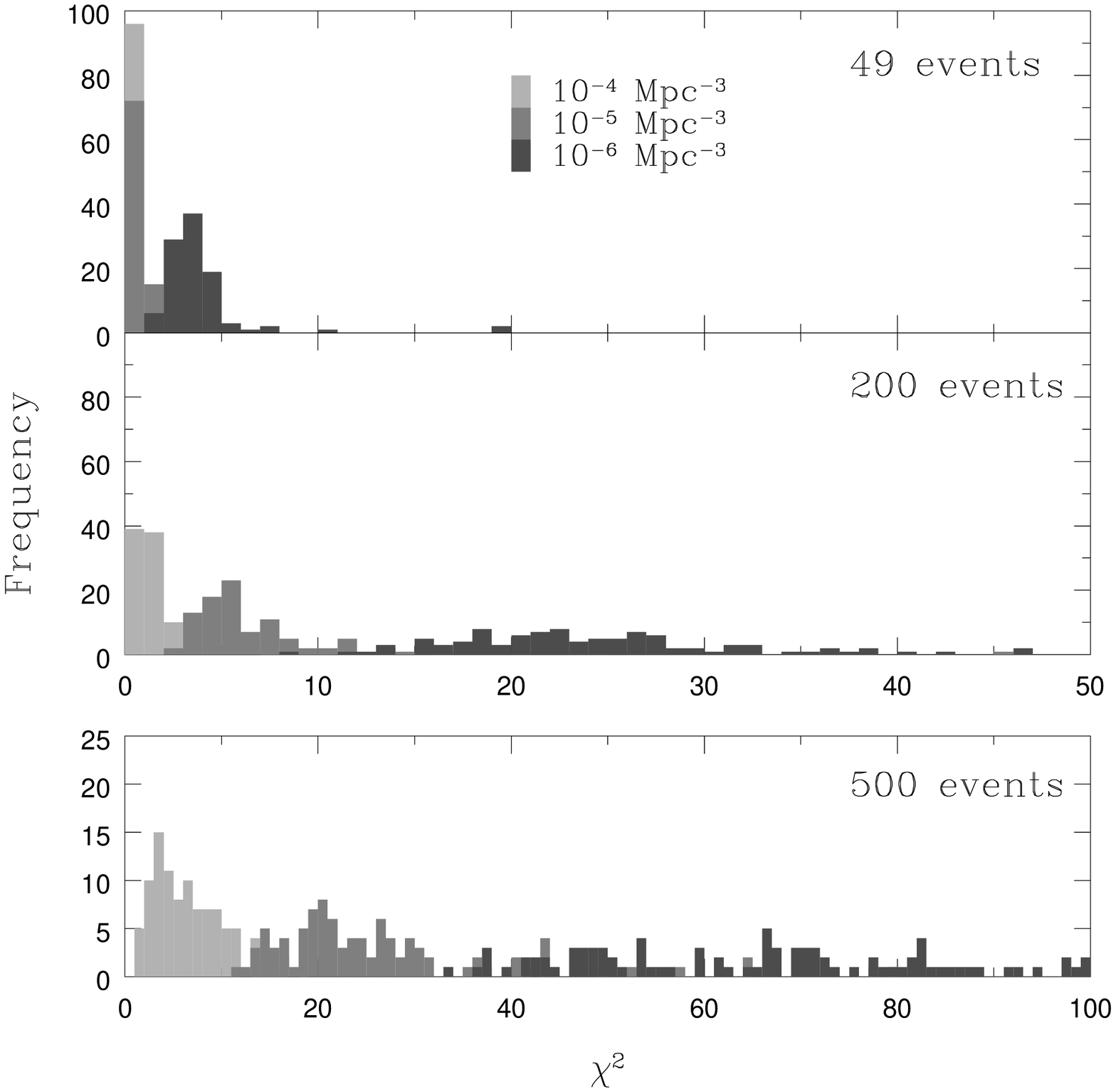}
\includegraphics[width=0.30\linewidth]{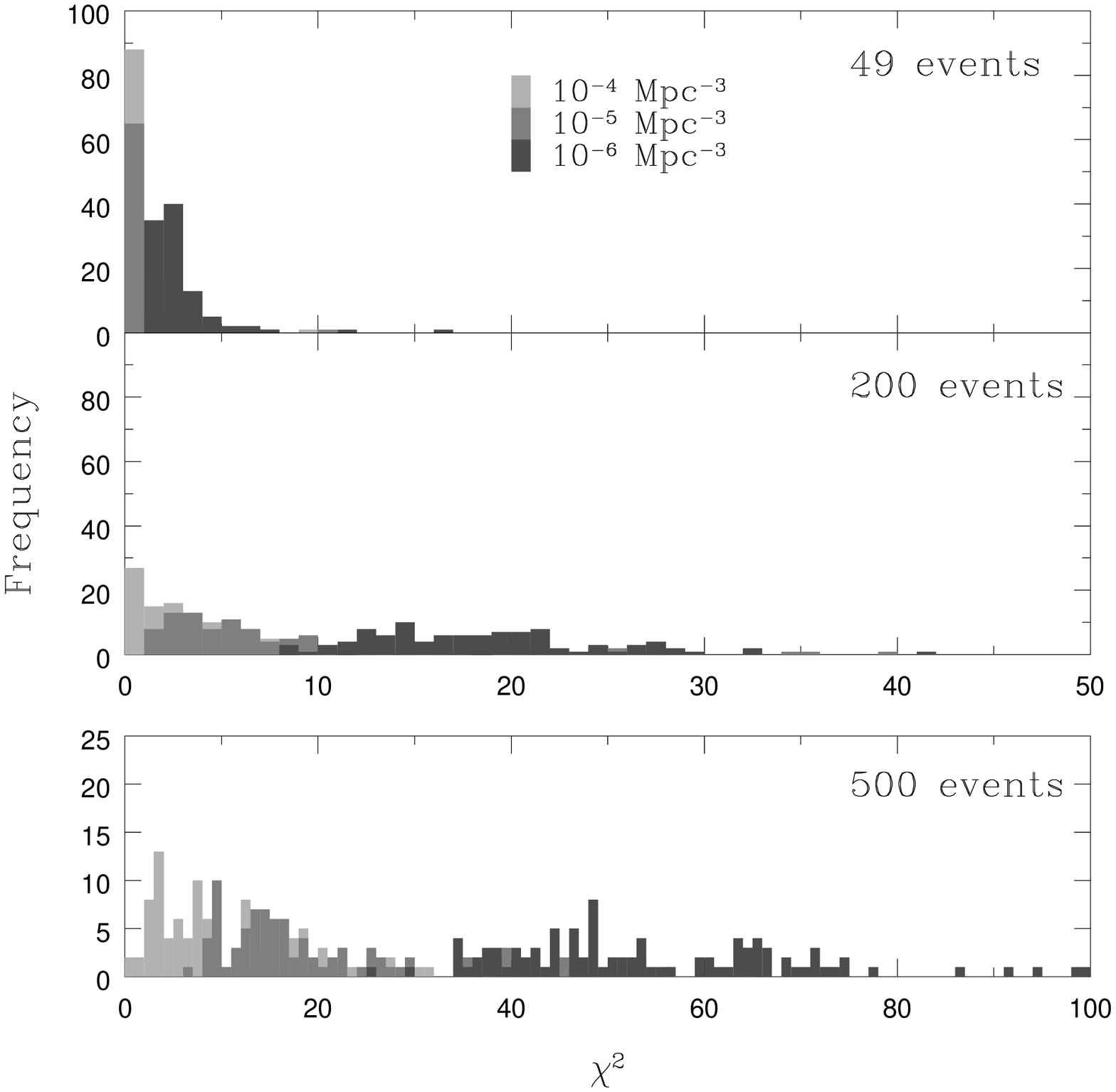}
\includegraphics[width=0.30\linewidth]{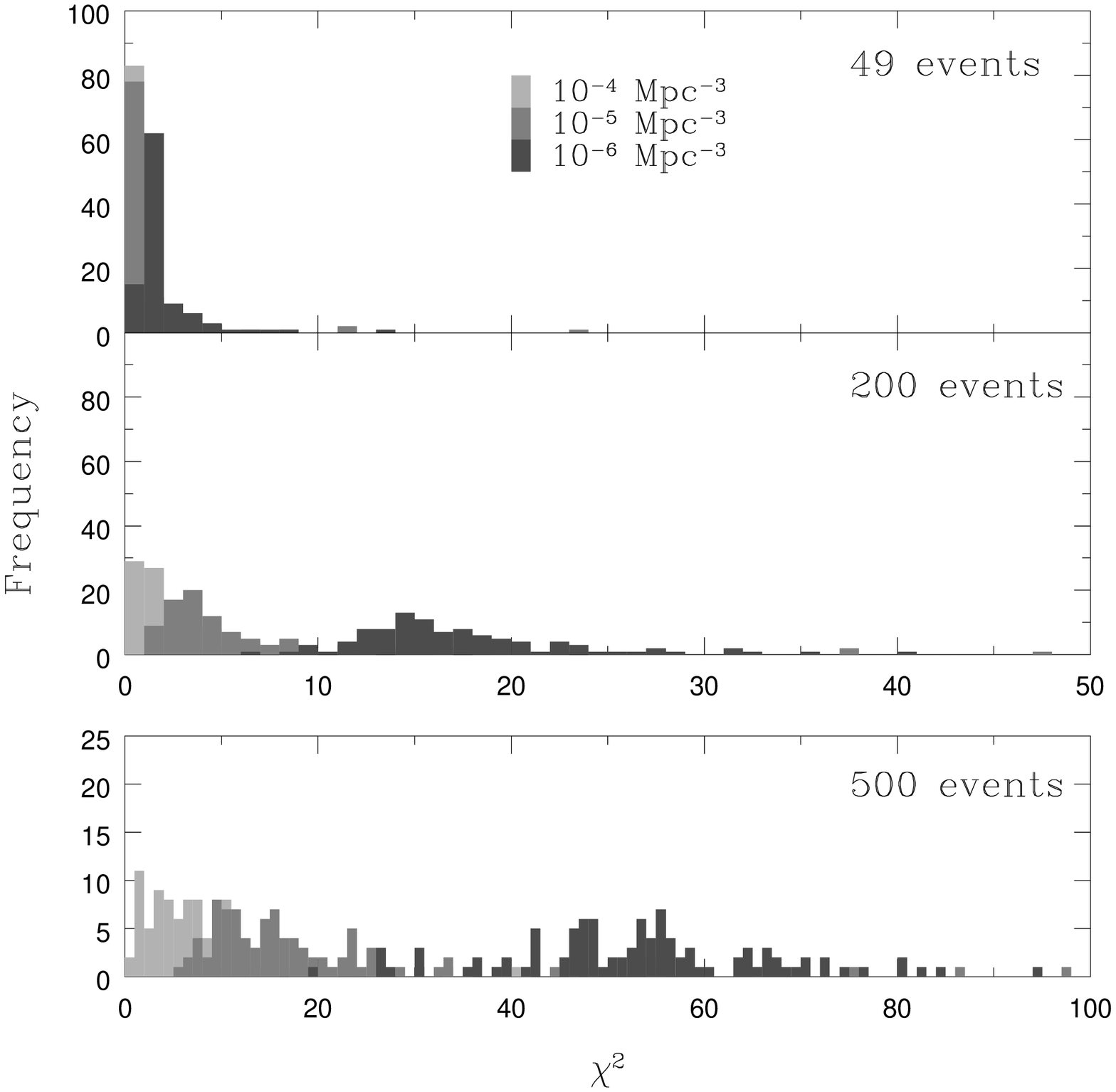}
\caption{Distributions of $\chi^2$s, 
calculated from the arrival protons above $4 \times 10^{19}~{\rm eV}$, 
at several strengths of the EGMF. 
The GMF is considered. 
The strengths of the EGMF are 0.0({\it left}), 0.1, ({\it middle}), and 
1.0$\mu{\rm G}$({\it right}). 
The numbers of events are set to be 49 events within 
$-10^{\circ} < \delta < 80^{\circ}$ ({\it upper}), 
200 events ({\it middle}), and 500 events ({\it lower}). 
within the southern hemisphere to emulate Auger.}
\label{fig3}
\end{center}
\end{figure*}

Figure. \ref{fig3} shows distributions of $\chi^2$ defined as 
\begin{equation}
\chi^2 \equiv \frac{1}{\theta_{\rm max}} 
\sum_{\theta=1^{\circ}}^{\theta=\theta_{\rm max}} 
\frac{\left[ N_{\rm sim}(\theta) - N_{\rm iso}(\theta) \right]^2}
{{\sigma_{\rm sim}(\theta_{\rm})}^2 + {\sigma_{\rm iso}(\theta)}^2}. 
\end{equation}
This value represents the goodness of the fitting. 
The upper panels show current status corresponding to the AGASA result. 
The three distributions with $10^{-4}$, $10^{-5}$ and $10^{-6}~{\rm Mpc}^{-3}$ 
are almost overlaped. 
The determination of the number density 
has large uncertainty. 

200 event observation allows us to discriminate 
$10^{-5}$ and $10^{-6}~{\rm Mpc}^{-3}$ with good accuracy 
since it can separate the distributions. 
This event number is probably comparable with current status of Auger. 
Detection of more events can divorce distributions 
with $10^{-4}$ and $10^{-5}~{\rm Mpc}^{-3}$. 
When 500 events are observed, 
the number densities of $10^{-4}$ and $10^{-5}~{\rm Mpc}^{-3}$ 
are perfectly separated 
if EGMF does not exist or is very weak.

\section{Conclusions}

In this study, we constrain number density of UHECR sources 
with the SSA in several strengths of EGMF and 
investigate future prospects for this approach. 
At the AGASA era, 
the source number density is $\sim 10^{-5}~{\rm Mpc}^{-3}$ 
with uncertainty of about an order of magnitude 
due to small number of observed events. 

That near future observations increasing observed event number 
can improve its uncertainty. 
200 event observation above $4 \times 10^{19}~{\rm eV}$ 
can distinguish $10^{-6}~{\rm Mpc}^{-3}$ from the more source density. 
This event number is consistent with the number observed by Auger until 
this summer! 
More event detection enables us to estimate the source number more accurately 
and, then, to be easy to compare it with that of known powerful objects. 
Number of events to discriminate between $10^{-4}$ 
and $10^{-5}~{\rm Mpc}^{-3}$ is dependent on the EGMF strength. 

Finally, we discuss results of another source model 
that power of each source is proportional to its luminosity, 
as discussed in \cite{takami06}. 
The latter model predicts $10^{-4} \sim 10^{-5}~{\rm Mpc}^{-3}$ 
at present. 
The source number density increases 
since dark sources are also counted, 
but hardly contribute the arrival cosmic rays. 
More observation can discriminate $10^{-3}~{\rm Mpc}^{-3}$ 
from less number densities. 
This model has an additional degree of freedom by luminosity, 
compared with the former model. 
This provides large dispersion to distribution of $\chi^2$. 
Therefore, $10^{-4}$ and $10^{-5}~{\rm Mpc}^{-3}$ cannot 
be distinguished even at 500 event observations. 

\section{Acknowledgements}

The work of H.T. is supported by Grants-in-Aid for JSPS Fellows. 
The work of K.S. is supported by Grants-in-Aid 
for Scientific Research provided by the Ministry of Education, 
Science and Culture of Japan 
through Research Grants S19104006.

\end{document}